\documentclass{emulateapj}

\shorttitle{Pal 5's Luminosity Function: Cluster vs.\ Tails}
\shortauthors{Koch et al.}

\begin{document}

\title{Mass Segregation in the Globular Cluster Palomar 5 and its Tidal
Tails} 

\author{Andreas Koch\altaffilmark{1,2}, 
Eva K.\ Grebel\altaffilmark{1,2}, 
Michael Odenkirchen\altaffilmark{2}, 
David Mart\'{\i}nez-Delgado\altaffilmark{2}, 
and John A.\ R.\ Caldwell\altaffilmark{2,3} }
\altaffiltext{1}{Astronomical Institute of the University of Basel,
Venusstrasse 7, CH-4102 Binningen, Switzerland }
\altaffiltext{2}{Max-Planck Institute for Astronomy, 
K\"onigstuhl 17, D-69117 Heidelberg, Germany }
\altaffiltext{3}{Space Telescope Science Institute, 3700 San Martin Drive,
Baltimore, MD 21218}

\begin{abstract} 
We present the stellar main sequence luminosity function (LF) of the
disrupted, low-mass, low-concentration globular cluster  Palomar 5 
and its well-defined tidal
tails, which emanate from the cluster as a result of its tidal
interaction with the Milky Way.  The results of our deep (B$\sim$24.5)
wide-field photometry unequivocally indicate that preferentially
fainter stars were removed from the cluster so that the LF of the
cluster's main body exhibits a significant degree of flattening
compared to other globular clusters.  There is clear evidence of 
mass segregation, which is reflected in a radial variation of the LFs.
The LF of the tidal tails is distinctly enhanced
with faint, low-mass stars.  Pal\,5 exhibits a binary main sequence,
and we estimate a photometric binary frequency of roughly 10\%.
Also the binaries show evidence of mass segregation with
more massive binary systems being more strongly concentrated
toward the cluster center.

\end{abstract}

\keywords{Galaxy: evolution --- Galaxy: halo --- globular clusters: general ---
globular clusters: individual (Palomar 5) --- stars: luminosity function}

\section{Introduction}

Globular clusters exhibit a range of different luminosity functions
and hence mass functions.  Since globular clusters tend to be very old
systems with ages $>10$ Gyr, main sequence luminosity functions sample
low-mass stars with masses typically below  0.8 M$_{\odot}$.  In this
regime the mass function begins to deviate from the canonical Salpeter
(1955) slope of $x = 1.35$.  For the {\em initial}\ mass function,
Kroupa (2001) derives an empirical composite power law with a slope $x
= 1.3 \pm 0.3$ for stars with masses from 0.5 to 1 M$_{\odot}$, but $x
= 0.3 \pm 0.5$ in the mass range of 0.08 to 0.5 M$_{\odot}$.
Comparing globular cluster {\em present-day}\ mass functions, Piotto
\& Zoccali (1999) find flat slopes with values of $-0.5 < x < 0.5$ for
stars with masses $< 0.7$ M$_{\odot}$.  

Piotto \& Zoccali (1999) point out that the global shape of globular
cluster luminosity functions is similar.  The slope is characterized by a
steep rise from brighter absolute magnitudes to $M_V \sim 10$ mag and
a drop thereafter.  This drop in the {\em luminosity} functions is
caused by the small number of stars over large magnitude bins at
fainter magnitudes (corresponding to masses $\le 0.2$~M$_{\odot}$),
whereas the {\em mass} functions continue to rise roughly
monotonically to 0.15 M$_{\odot}$ (Chabrier \& M\'era 1997).
Luminosity functions have the advantage that they are directly
measurable, while the conversion from luminosity functions to mass
functions is still plagued by uncertainties in the mass-luminosity
relation (e.g., Chabrier \& M\'era 1997).

Some globular clusters have flatter present-day mass functions than
others.  In particular, flatter slopes are found in clusters with
smaller half-light radii, at smaller Galactocentric distances, and
with higher destruction rates (Piotto \& Zoccali 1999).  The weak
dependence on metallicity, on the other hand, seems to be primarily an
effect of Galactocentric distances (Chabrier \& M\'era 1997).  These
correlations are believed to be primarily caused by internal dynamical
evolution coupled with external tidal stripping.  Two-body encounters
between individual stars within globular clusters increase the
velocities of lower-mass stars, which in turn increases their average
distances from the cluster center as compared to higher-mass stars
(e.g., Aguilar, Hut, \& Ostriker 1988; Oh \& Lin 1992; King, Sosin, \&
Cool 1995; Murray \& Lin 1996; Meylan \& Heggie 1997).  Hence these
low-mass stars can more easily be removed by Galactic tides.  The
stronger gravitational field at smaller Galactocentric radii leads to
more extensive stripping, while clusters in the distant halo should be
able to hold on to their low-mass stars more easily.  To first order,
this simple scenario can account for the above described correlations.  

Additional properties such as the stellar density within a cluster or
concentration as well as the eccentricity and orientation of cluster
orbits also play a role and add to the observed scatter in the
correlations.   In particular, in clusters with low central
concentration the stellar encounter rate is reduced.  As noted by,
e.g., King et al.\ (1995), the magnitude of mass segregation depends
on the depth of the potential well or on the degree of central
concentration.  Mass segregation, where present, varies as a function
of distance from the cluster center.  Mass segregation is most
pronounced in the cores of globular clusters, but tends to be much
less noticeable at larger distances from the cluster center.
Measurements of the luminosity function at a cluster's half-mass or
half-light radius typically show little evidence for segregation and
dynamical modification (e.g., Lee, Fahlman, \& Richer 1991; Chabrier
\& M\'era 1997; Piotto \& Zoccali 1999; Paresce \& De Marchi 2000).  

In this paper we present a study of the luminosity function of 
Palomar 5, an unusual halo globular cluster.  Pal\,5 is a very 
sparse, extended, low-concentration and
low-mass cluster on a highly eccentric orbit (see Tab.\ 1 
for a list of properties of Pal\,5).  

\begin{table}[h]
\begin{center}
\caption{Pal 5's main parameters}
\begin{tabular}{l r c}
\colrule
\colrule
Parameter & Value & Reference \\
\colrule
$\alpha (J2000)$ & 15 16 04.6 & a \\
$\delta (J2000)$ & 00 07 15.6 & a \\
Galactocentric distance & 18.6\,kpc & b \\
Orbital eccentricity $e$ & 0.46 & c \\
Core radius r$_c$ & 3$\farcm$6 (24.3\,pc) & a  \\
Tidal radius r$_t$ & 16$\farcm$1 (109\,pc)& a \\
Concentration c & 0.66 & a  \\
Mass & 5$\cdot$10$^3\,M_{\sun}$ & a\\
Radial velocity (heliocentric) & $-$58.7\,km\,s$^{-1}$ & a \\
Velocity dispersion & 1.1\,km\,s$^{-1}$ & a \\
Metallicity [Fe/H] & $-$1.43  & b \\
Relaxation time \symbol{64} half-mass radius & 7.8\,Gyr  & b \\
Age & 11.5\,Gyr & e\\
\colrule
\end{tabular}
\tablecomments{The main parameters characterizing the globular cluster Palomar 5. The values are taken from $^a$Odenkirchen et al. (2002), $^b$Harris (1996, 2003),
$^c$Odenkirchen et al. (2001), $^d$Odenkirchen at al. (2003), $^e$Martell et al. (2002).}
\end{center}
\end{table}

Recently well-defined, narrow
tidal tails emanating from Pal\,5 were discovered in wide-field
photometric data from the Sloan Digital Sky Survey (SDSS)
(Odenkirchen et al.\ 2001).  These tails have since been traced over
more than 10$\degr$ across the sky (Odenkirchen et al.\ 2003).  The mass 
contained in the tails exceeds the current mass of the cluster by at least
a factor of 1.2; possibly by more since the tails probably extend over a much
larger area than surveyed to date.  With a total present-day mass of only about
5000\,$M_{\odot}$, Pal 5's main body has suffered severe mass loss.
An estimate of the loss rate yields 5\,$M_{\odot}\,$Myr$^{-1}$, which 
implies that at least 90\% of its initial stellar content have been lost 
(Odenkirchen et al. 2003).  Following the N-body simulations of Dehnen
et al.\ (2004), the original mass of Pal\,5 is estimated to have been
$\sim 70,000$~M$_{\odot}$ if the mass loss rate of Pal\,5 had been 
constant over its lifetime.

One would expect that a low-concentration cluster should have experienced
little internal dynamical evolution if its low concentration was typical
for most of its evolution.  Consequently, one would also expect that its
luminosity function should show nothing out of the ordinary, and that
dynamical mass segregation should be non-existent.  

The most recent detailed analysis of Pal\,5's main sequence is based 
on observations with the Hubble Space Telescope (HST) (Grillmair \& Smith 2001,
hereafter GS01).  GS01 found a significant flattening at the faint end of
the main sequence luminosity function of Pal\,5 compared to other 
globular clusters.
The limiting magnitude of the HST photometry is V$<$27.5, providing the 
deepest existing color-magnitude diagram until now.  The HST data do not
show evidence for spatial variations in the luminosity function, but also
only cover a small field of view (two fields of $\approx 2\farcm 4$ squared 
each).  GS01 interpret this as lack of mass segregation within the cluster
core (both of their overlapping fields lie within Pal\,5's core radius).  

The detection of tidal tails around Pal\,5 prompted us to revisit the 
unusually flat luminosity function (LF) of the cluster itself and to extend
this kind of study to its tails.  The deficiency in low-mass stars observed
in the center of Pal\,5 raises the question whether this cluster might have
had an unusual luminosity and mass function to begin with, or whether
other effects might have played a role during its evolution, e.g.,
internal (relaxation) and external (disk shocking) dynamical effects.  
A recent analysis of the LF both of Pal 5's central region and its tidal 
tails, based on SDSS data, revealed no variation of the LF between these 
two areas  (Odenkirchen et al. 2003). However, these LFs only comprised 
the giant branch down to the upper main sequence to approximately 1\,mag 
below the turn-off (i.e., i$^{\ast}\la 21.8$). Effects of mass loss and 
mass segregation, if present, should become more pronounced at fainter
magnitudes along the main sequence, where luminosity reflects mass,
but these magnitudes are beyond the limits of SDSS data.

In the present study we measure LFs for different 
regions of the faint globular cluster Pal 5 and its tails. 
While our photometry is not as deep as the HST data, our wide-field 
observations cover a much larger area, providing us with good number
statistics several magnitudes below the main sequence.

This paper is organized as follows:  Section 2 describes our
observations and the photometric reduction steps. Color-magnitude
diagrams and the derivation of the LFs are described
in Section 3. In Section 4, the resulting LFs for different regions of
Pal 5 are presented. Section 5 presents the mass functions, which were
derived from the observed LFs, whereas Section 6 deals with the cluster's
photometric binary component.  A recapitulatory comparison of Pal 5 with other
globular clusters' LFs is given in Section 7. Finally, Section 8
summarizes the results.

\section{Data and Reduction}

Imaging data of the cluster Palomar 5 and several regions located in its
tidal tail were obtained in two observing runs using two different 
wide-field instruments: The Wide Field Imager camera (WFI) at the ESO/MPG
2.2\,m telescope at the European Southern Observatory at La Silla in
Chile, and the Wide Field Camera (WFC) at the 2.5\,m Isaac Newton
Telescope (INT) at the Observatorio de Roque de los Muchachos (La
Palma, Spain).  These instruments both cover a large field of view, which is
vital for efficient observations of portions of the very extended,
low-density tidal tail.  Odenkirchen et al. (2003) have shown that the
stellar surface density in the tails does not exceed
0.2\,arcmin$^{-2}$, a very low density indeed when compared to the field
star density of 0.16\,arcmin$^{-2}$ in the surroundings.  Furthermore,
the cluster itself has a large angular extent on the sky: Its apparent tidal
radius is approximately 16$\arcmin$; again making wide-field imagers
the instruments of choice.  Our follow-up observations with these
instruments aim at an analysis of the cluster's LF 
down to magnitudes below the SDSS detection limit in order to search
for possible spatial variations and mass segregation effects.

\subsection{WFI observations}

The observations were performed with the WFI during photometric
conditions (Table 2).  The WFI camera consists of a mosaic of eight
CCDs, each comprising 2046$\times$4098 pixels with a pixel scale of $0\farcs
238$\,pixel$^{-1}$. The field of view of the WFI is $34\arcmin \times
34\arcmin$. The individual CCDs are separated by gaps with a width of
$23 \arcsec$.  Apart from the central part of the cluster (labelled
field F1) we targeted a second field in the trailing tail (F2).
Furthermore we observed a comparison field (F3), located well away
from the cluster (1$\fdg$5) and the tails, in order to estimate the
characteristics of field stars in this region.  Observations were
taken in the WFI's V and R filters, which are similar, but not
identical to the Johnson-Cousins V,R$_C$ bands (Johnson \& Morgan
1953; Cousins 1978; Girardi et al.\ 2002).  Each of the fields was 
observed five times
in each filter.  The exposures were dithered against each other by
approximately 15$\arcsec$ in the vertical and horizontal direction in
order to cover the gaps between the single WFI CCDs.  The seeing during
both nights did not exceed 1$\farcs$4, and the airmass was 1.2 on
average.

\subsection{INT observations}

In addition, observations of Palomar 5 and its tidal tails were
carried out using the WFC in the prime focus of the INT.
The WFC contains four $2048\times 4096$ pixels EEV CCDs. The pixel scale is
$0\farcs 333$, which provides a total field of $35\arcmin \times
35\arcmin$.  In this run, four fields in and around Palomar 5 were
observed: One targeting the cluster's center (labelled A), two
targeting the density clumps located in the northern ({\em trailing})
tail (B) and the southern ({\em leading}) tail (C), and finally a
control field (D), $1\fdg 4$ away from the central field.
Additionally bias and twilight flatfield exposures were obtained in
each night.  The observations were obtained using a Harris B and an
SDSS r filter (for filter definitions
see Gunn et al.\ 1998; for a comparison with Johnson-Cousins filters
see Grebel 2001) under photometric
conditions.
The seeing ranged from $1\arcsec$ to $1\farcs 2$ in both bands. 

An overview of our observations is presented in Table 2.  The location of
our fields is depicted in Fig.\ 1.

\begin{figure}[h]
\includegraphics[angle=0,clip,width=1\hsize]{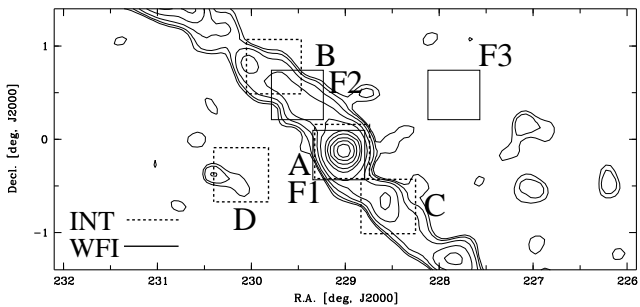}
\caption{Observed fields, overlaid on a contour map
of the stellar surface density of Pal 5 from SDSS data (Odenkirchen et
al.\ 2003).}
\end{figure}

\begin{table*}[t]
\begin{center}
\caption{Summary of the observations.\label{tbl-2}}
\begin{tabular}{l c c c c c }
\colrule
\colrule
Field & $\alpha$ & $\delta$ & Instrument & Exp. time & Date \\
      & (J2000)  & (J2000)  & \& Filter  & [s]       &      \\
\colrule
Pal 5 center  (F1)   & 15 16 21 & $-$00 10 48 & WFI R & 5$\times$600 & 2001 May 18 \\
		     &		   &		  & WFI V & 5$\times$900 & 2001 May 18 \\
Northern tail (F2)   & 15 18 10 & $+$00 31 48 & WFI R & 5$\times$600 & 2001 May 17 \\
		     &		   &		  & WFI V & 5$\times$900 & 2001 May 17 \\
Control field 1 (F3) & 15 11 12 & $+$00 31 48 & WFI R & 5$\times$600 & 2001 May 18 \\
		     &		   &		  & WFI V & 5$\times$900 & 2001 May 17 \\
\colrule
Pal 5 center  (A)    & 15 16 05 & $-$00 06 36 & INT B & 2$\times$1000& 2001 Jun 23 \\  
		     &		   &	          & INT r & 3$\times$900 & 2001 Jun 23 \\
Northern Tail (B)    & 15 19 00 & $+$00 48 00 & INT B & 2$\times$1000& 2001 Jun 24\\  
		     &		   &		  & INT r & 3$\times$900 & 2001 Jun 24 \\
Southern Tail (C)    & 15 14 07 & $-$00 42 00 & INT B & 2$\times$1000& 2001 Jun 25\\  
		     &		   &		  & INT r & 3$\times$900 & 2001 Jun 25\\
Control field 2 (D)  & 15 20 24 & $-$00 22 12 & INT B & 2$\times$1000& 2001 Jun 26 \\  
		     &		   &		  & INT r & 3$\times$900 & 2001 Jun 26 \\
\colrule
\end{tabular}
\end{center}
\end{table*}

\subsection{Photometric reduction}

\subsubsection{Basic reduction}

The raw data files were split into four (INT WFC) or eight (WFI)
single images, respectively, each corresponding to one individual CCD
chip. Thus, during all of the subsequent reduction steps, each of the
chips was treated separately. The standard reduction steps were
carried out using the IRAF package\footnote{IRAF is distributed by 
the National Optical Astronomy Observatories, which are operated by 
the Association of Universities for Research in Astronomy, Inc., 
under cooperative agreement with the National Science Foundation.}.  
Readout bias was removed to
first order by subtracting a fit of the overscan region from the
frames.  Any residual bias was subtracted using a mean over
approximately 30 bias frames.  Flatfield calibration was carried out
using the qualitatively best of the observed twilight flats.  Finally,
bad pixels and columns were masked out.  Neither dark current nor
fringing causes any considerable effect in either our WFI and INT
observations.

\subsubsection{WFI Photometry}

Details of the photometric processing and calibration of the WFI
frames are described in a separate paper (Koch et al.\ 2004).
Basically, the reduced frames were processed using the DoPHOT package
(Schechter, Mateo \& Saha 1993).  The actual photometry was calculated
by fitting an analytic point spread function to all detected objects.
As we are only interested in stars for our following work, we rejected
all non-pointlike objects from our photometry list.  Afterwards V and
R output data were matched against each other regarding position on
the images, retaining only objects detected in both filters.  Since
the WFI does not produce entirely spatially homogeneous photometry due
to central light concentration (e.g., Manfroid, Selman \& Jones 2001;
Koch et al.\ 2004), we applied photometric correction terms to remove
large-scale spatial gradients.  These correction terms were derived by
comparing our instrumental magnitudes to the well-calibrated sample of
stars from the SDSS Early Data Release (EDR, Stoughton et al.\ 2002),
which coincide with our fields. 
For this purpose we defined equations of the kind
\begin{eqnarray}
R^{\ast}&\,=\,&r+\alpha_R\,(g-r)\,+\,\beta\,(r-i)\,+\,c_R, \\
V^{\ast}&\,=\,&g+\alpha_V\,(g-r)\,+\,c_V, 
\end{eqnarray} 
with g, r, and i being SDSS magnitudes. WFI instrumental magnitudes are denoted
by an asterisk.
The remaining residuals after applying the transformation equations (1, 2), 
$\varepsilon$,
showed strong spatial dependence and were fitted by a second order
model ($\varepsilon=\varepsilon(x,y)$), which then was subtracted from the instrumental magnitudes.
Finally, the transformation to standard Johnson magnitudes was
obtained from Table 7 of Smith et al.\ (2002), where sets of transformation equations 
between the SDSS and Johnson standard systems are provided. 
Accounting for zero-point offsets
between our data and the tabulated values of Smith et al.\ (2002), we get
\begin{eqnarray}
R&\,=\,&R^{\ast}\,-\,c_R\,-\,\varepsilon_R(x,y) \\
V&\,=\,&V^{\ast}\,-\,c_V\,-\,\varepsilon_V(x,y).
\end{eqnarray}
The actual values of the coefficients
($\alpha,\,\beta$) and zeropoints ($c$) are tabulated for each of the
eight WFI CCDs in Koch et al. (2004), together with the spatial model for the 
variations in $\varepsilon(x,y)$.

\subsubsection{INT Photometry}

DAOPHOT and ALLSTAR (Stetson 1987, 1994) were used to obtain the stellar
photometry.  To avoid contamination with
background galaxies, we imposed the following constraints on crucial
ALLSTAR parameters: CHI$\,<\,$2 and $-1 <\,$SHARP$\,< 1$.  This
basically rejects extended objects and merely leaves point sources
with stellar point spread functions.  
To transform these data to a
standard photometric system we proceeded analoguously to
Sect. 2.3.2, i.e., by comparing our set of
instrumental magnitudes to the SDSS EDR photometry, which may be
regarded as a set of local tertiary standards and which overlap fully
with our observed fields.
INT's r-filter is identical to the SDSS r-definition, hence a linear 
fit yielded the transformation to the standard SDSS sytem for each chip. 

The B-magnitudes were converted into Johnson magnitudes similar to eqs. (1,3), where B is written as a linear
combination of u, g and r, again with final account 
for offsets when compared to Smith et al. (2002).

The formal transformation errors from the fit dispersion are smaller than 0.01\,mag for each chip and likewise
for the zero-point error from the comparison with Smith et al. (2002) thus placing an estimate of approximately 
0.02\,mag on this error source.  
Additionally, DAOPHOT provides a dispersion of the PSF fitting for each star. These uncertainties $\sigma$ are
(0.005, 0.008, 0.015, 0.035, 0.1) on average at B\,=\,(20, 21, 22, 23, 24)\,mag.   
Putting all these errors together, the total photometric error of our data can be estimated to be no larger than
0.15\,mag at the very faintest magnitude bin that is going to be used in the subsequent analyses.
Due to, e.g., the stronger spatial variations in the WFI photometry the respective uncertainties are larger, reaching approximately 0.2\,mag
at the faint end.

\section{Color-magnitude diagrams and number counts}
 
\subsection{Color-magnitude diagrams}

The diagrams presented in Figures 2 (WFI data) and 3 (INT data)
contain data of all stellar objects found in the observed fields. 
The INT data reach magnitudes fainter by approximately 1\,mag than the
WFI data. The main sequence seen in the INT data is significantly
narrower than that derived from the WFI observations, which we ascribe
to the better seeing, larger color baseline,
fainter magnitude limits, and hence better
signal to noise of the INT data.  Despite the limitations of the WFI
data quality we chose to keep them for further analysis.  These data
do not only provide a second control field for estimating field star
contamination, but also cover a different part of the tidal tails than
the INT data. 

\begin{figure}[h]
\includegraphics[angle=0,clip,width=9cm]{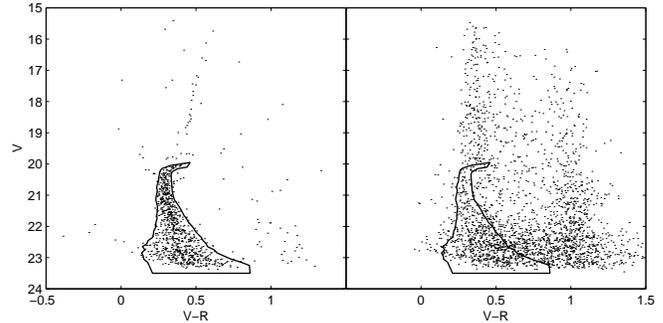}
\caption{Color-magnitude diagrams (CMDs) of different regions
observed with the WFI. The {\em left} panel shows stars within one core radius
(3$\farcm$6, field F1), whereas the CMD of field F2 in the northern
tail is shown in the {\em right} panel. Additionally
shown is the 2$\sigma$-envelope around the observed, averaged main
sequence from the central part of the cluster that will be used for an
assessment of the field contamination in Section 3.3.}
\end{figure}

\begin{figure*}[t]
\includegraphics[angle=0,clip,width=1\hsize]{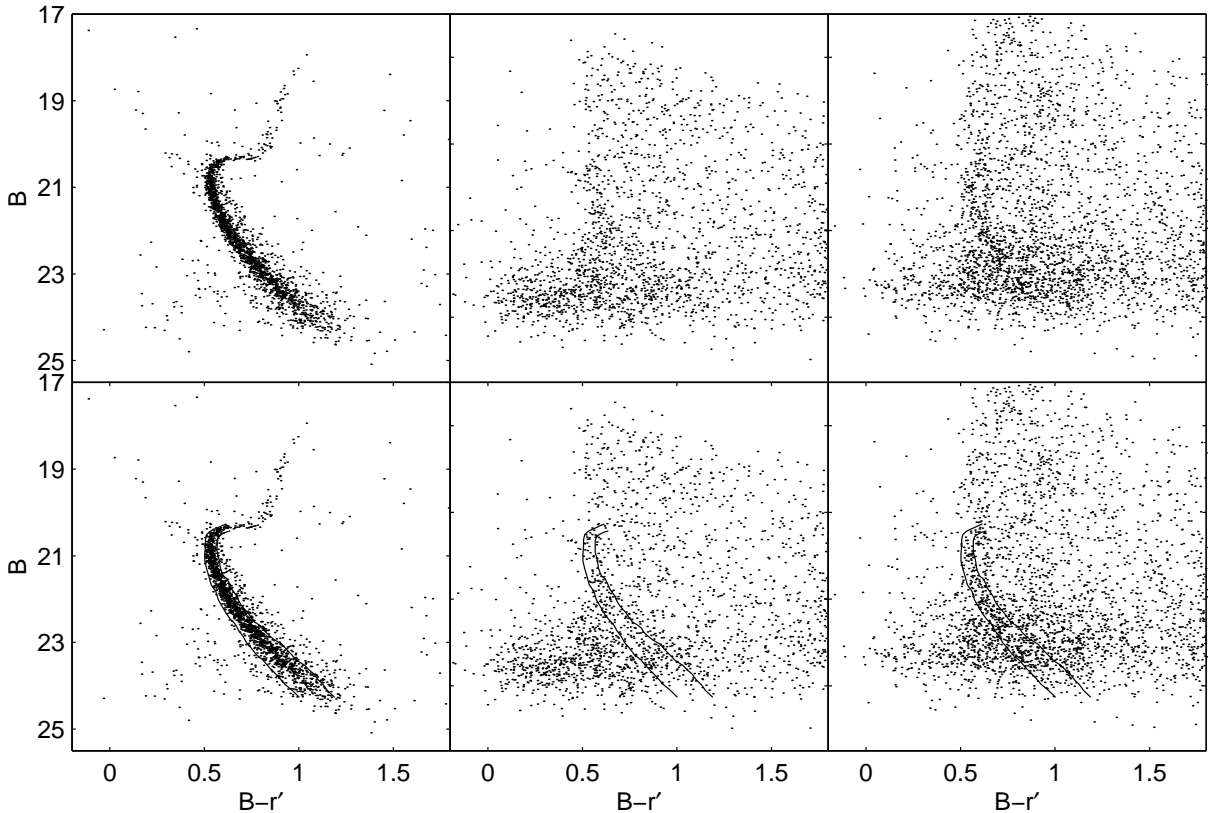}
\caption[Koch.fig3.eps]{Same as Figure 2, but for data taken with the INT:
{\em Left} panels show the central region, {\em middle} panels display stars 
from field B in the northern tail, whereas the CMD for the southern tail is depicted
on the {\em right} hand side. The lower panel additionally shows the 2$\sigma$-envelope
of the main sequence.}
\end{figure*}

The basic features of the color-magnitude diagrams (CMDs) are as follows.  

\begin{enumerate}
\renewcommand{\labelenumi}{(\arabic{enumi})} 

\item Cluster population
(left panels): Within the core radius of Pal\,5 of 
3$\farcm$6, the contamination with
field stars is rather small ($\sim  4$\%). Thus the CMD contains mainly
stars belonging to 
the cluster population.  Characteristic features are the main sequence
of unevolved stars below V$\approx$20\,mag, the turnoff-point at
V$\approx$20.8 (Smith et al. 1986) and the narrow subgiant branch.
The red giant branch is visible, although sparsely populated (see also
Odenkirchen et al.\ 2001).  Obvious horizontal branch stars are not
present in our CMDs.  Redward of the main sequence, 
there is a binary sequence visible in the INT data, which
will be investigated in more detail in Sect. 6.

\item Tail population (middle/right panels): As the sparsely populated
regions in the tails contain a large fraction of field stars, the 
signature of the evaporated cluster stars does not stand out
clearly.  However, the main sequence within the southern tail is more
distinct than that of the northern tail, which may be due to the
closer distance of field C to the cluster (at 3.3 tidal radii compared
to 4.3 radii for field B).  As was shown by Odenkirchen et al.\
(2002), the density declines with increasing distance from the cluster's
main body.  

\end{enumerate}

\subsection{Incompleteness}

In order to account for incompleteness effects in our 
data we performed artificial star experiments. The magnitudes of the stars
added to the WFI 
frames were chosen to cover a range of 20.25 to
23.25\,mag in steps of 0.5\,mag, which we defined as central value for
each magnitude interval [V$-0.25$,V$+0.25$]. One star-added image was
produced for each magnitude bin for each initial image and again for
each of the eight WFI chips. The artificial stars were spatially
distributed on a predefined grid such that their total number amounts
approximately 10\% of the stellar objects found by DoPHOT on the
initial frames. This procedure was chosen to avoid introducing
artificial crowding (Andreuzzi et al. 2001). 

Colors and magnitudes chosen for 9000 artificial stars on the INT images 
were obtained from an isochrone resembling the observed  main sequence 
(11.2\,Gyr, [Fe/H]=$-$1.30, see Section 5)
and distributed on a grid across the entire images. Afterwards 
all star-added frames were processed through the DoPHOT and DAOPHOT
routines using  exactly the same parameter
set as in the original reductions. 
An artificial star was defined as
recovered if it met two conditions: Firstly, its position had to be
identical (to within the fitting uncertainties) to the input grid
position. And secondly the output magnitude was required to still lie
within the input magnitude bin. 

We define the limiting magnitude as the
magnitude where  
the incompleteness fraction $\chi$, defined as the number ratio of stars recovered to those injected,   
drops below 50\% and find V$_{lim}$\,=\,23\,mag for our WFI data and 
a limiting B-band magnitude of B$_{lim}\approx 24$\,mag for the INT frames.
The incompleteness fraction versus B- and V-magnitude
(INT and WFI, respectively) is presented in Fig.\ 4. 
\begin{figure}[h]
\begin{center}
\includegraphics[angle=0,clip,width=8cm]{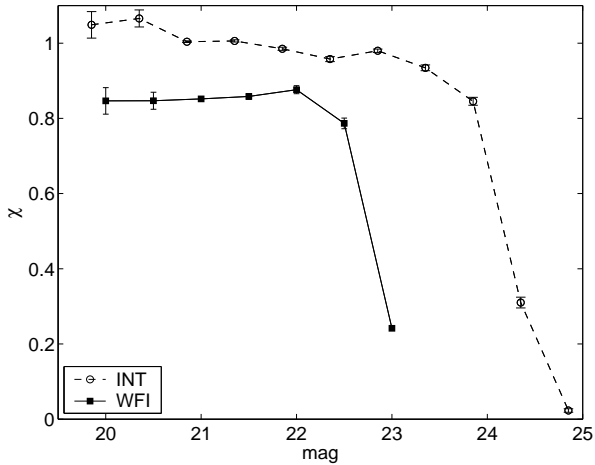}
\end{center}
\caption{Photometric completeness fraction $\chi$
for the WFI frames versus V-magnitude for one sample frame (solid
line). The dashed line represents values for $\chi$ from
our artificial star experiments on the INT data.}
\end{figure}

\subsection{Field contamination}

In order to count cluster main-sequence star candidates, it
is necessary to statistically remove field star contamination. 
This was done in two steps.  First, the mean locus of the main
sequence was calculated by averaging the photometric data of stars
within the cluster's core radius (left panels of Figs. 2,3).  Stars
obviously not belonging to the cluster (e.g., thick disk stars at
B$-$r$\,\ga$\,1.3) were excluded from the averaging process.
Afterwards an envelope comprising two standard deviations (2$\sigma$) of this averaged main
sequence fiducial was calculated for each magnitude bin.  The bin size for this
process was chosen to be 0.1\,mag. Finally, this 2$\sigma$-envelope was
smoothed in order to encompass a maximum number of obvious cluster
members.  This 2$\sigma$-method for defining the cluster in
color-magnitude space and transferring it to other fields provides a
compromise between a complete estimate of the cluster population and
minimizing the field contamination.
\begin{table*}[t] 
\begin{center} 
\caption{Field star contribution}
\begin{tabular}{cccccccc} 
\colrule
\colrule 
& \multicolumn{2}{c}{Central 3$\farcm$6} &
\multicolumn{2}{c}{Northern Tail}& Southern Tail
&\multicolumn{2}{c}{Comparison field}\\ 
Field $i$
& INT (A) & WFI (F1) & INT (B)  & WFI (F2) & INT(C) & INT (D)  & WFI
(F3) \\ 
\colrule 
N($i$)                                 &  1701 & 669      &   525   & 615     & 451 &   170       & 869     \\
$\frac{A_i}{A_{Field}}$                &  0.040      & 0.036    & 1.013 & 0.47    & 1.013  &    1        &  1      \\
N($Field$)$\cdot\frac{A_i}{A_{Field}}$ &     7       & 31       & 172   & 408     & 172 &   170       & 869     \\ 
\colrule
\end{tabular} 
\end{center} 
\tablecomments{Total number N of stars
(completeness corrected) within the smoothed 2$\sigma$-envelope for
each analyzed region and ratio of the respective area $A_i$ and that
of the field star observations, $A_{Field}$.  The last row gives the
number of statistically expected field stars.  The central 3$\farcm$6
encompass stars within Pal 5's  core radius.} 
\end{table*}

However, we will inevitably enclose field stars located within
2$\sigma$-limits of the main sequence.  To correct for this residual
contamination, observations of two comparison fields (F3 and D) were
taken.  The number of stars in the respective magnitude
bin in the comparison fields was subtracted from the number of stars in
the science fields, weighted by the ratio of areas covered.  The
fractional importance of contaminants is almost negligible if one
restricts the data to the region enclosed within the core radius
(4\%)\footnote{The gaps within the camera mosaics were taken into
account in calculating these ratios.}, but it accounts for a considerable
fraction of stars the farther one proceeds outward from the cluster center
(where the area weights are $\sim$50\%, 100\% respectively -- see Table 3).

To get an estimate for the magnitude of field contamination in each of the
observed regions (labelled $i$), Table 3 gives an overview over the
total number of stars N($i$) enclosed within the 2$\sigma$-envelope
down to the completeness limit and the ratio of the covered areas ($A$),
which -- combined with the density of cluster stars -- is a
measure of the fractional contribution of field stars. Also listed is the
number of field stars that is statistically expected in each region,
which is given by the number N($Field$)$\cdot\frac{A_i}{A_{Field}}$.

\section{The luminosity functions} 

We derive the 
LF from number counts in field $i$ in bins with a size of 0.5\,mag. 
Regarding this bin size, the small photometric uncertainties (see Sect. 2.3.3) do not introduce larger
errors in the final LFs due to bin migration. Such an effect would only 
affect the data at fainter magnitudes, where the incompleteness already becomes
severe. 
Accounting for incompleteness and field contamination, we define
the LF in the B band as

\begin{equation}
LF_i(B)\,=\,\frac{N_i(B)}{\chi_i(B)}\,-\,\frac{A_i}{A_{Field}}\,\frac{N_{Field}(B)}{\chi_{Field}(B)}
\end{equation} 

and likewise for the V band. 
Here $N$ is the number of stars recovered per bin and $A$
is the corresponding area of the field. As the only uncertainties we
assume purely statistical errors in the number counts and the 
incompleteness estimates. The errors in the counts are
assumed to follow a Poisson distribution, whereas uncertainties in the determination of $\chi$ 
are derived from a binomial distribution (see Bolte 1989; Andreuzzi et al.\ 
2004).  Conversion from apparent to
absolute magnitudes was obtained using a distance modulus of
(m$-$M)$_V=\,$16.9 and a reddening of E(B$-$V)\,=\,0.03 (Harris 2003).

Our completeness-corrected, field-subtracted stellar main sequence LFs
both for stars within Pal 5's core radius and in the tidal tails are
listed in Tables 4 (WFI) and 5 (INT). These LFs
are illustrated in Figs. 5 and 6.  

\begin{table}[h]
\begin{center}
\caption{WFI Luminosity Functions}
\begin{tabular}{ccrcr}
\colrule
\colrule
& \multicolumn{2}{c}{Cluster Center} & \multicolumn{2}{c}{Tidal Tail}\\
V & N & LF & N & LF\\
\colrule
20.0 &  19  & 21.6  & 14  & 6.4   \\
20.5 &  65  & 76.1  & 21  & 13.2  \\
21.0 &  103 & 121.2 & 31  & 22.6  \\
21.5 &  114 & 132.8 & 53  & 30.8  \\
22.0 &  120 & 128.0 & 119 & 62.7  \\
22.5 &  165 & 153.8 & 238 & 72.8  \\
23.0 &  246 & 153.9 &     &       \\
\colrule
\end{tabular}
\end{center}
\tablecomments{Absolute number counts N per magnitude bin [V,V+dV] and field subtracted, completeness corrected LF
from the WFI data -- both for stars within Pal 5's core radius (3$\farcm$6) and for a clump in the the tidal tail (Field F2).}
\end{table}

\begin{figure}[h]
\includegraphics[angle=0,clip,width=8.6cm]{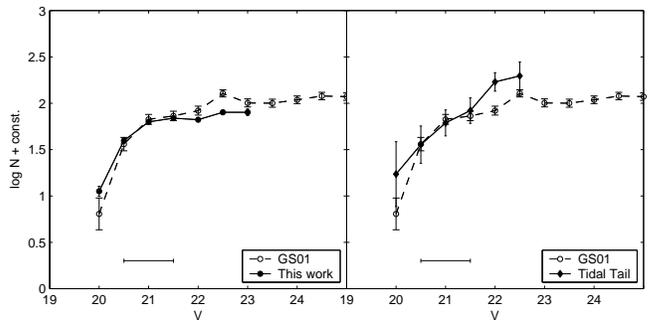}
\caption{{\em Left panel}: Completeness and field star
corrected WFI luminosity function (LF) in [V,V+0.5] 
for radii $r<r_c$.  Data from GS01 are shown as a dashed
curve. These comprise both their {\em core} and {\em
off-center} field.  The solid line at the bottom indicates the magnitude
interval used for normalization of the LFs. The  {\em
right} panel displays our results for the northern tidal
tail, also compared to the central LF from GS01. Errorbars derive from 
Poisson errors in the number counts.} 
\end{figure}

\begin{figure*}[t]
\includegraphics[angle=0,clip,width=1\hsize]{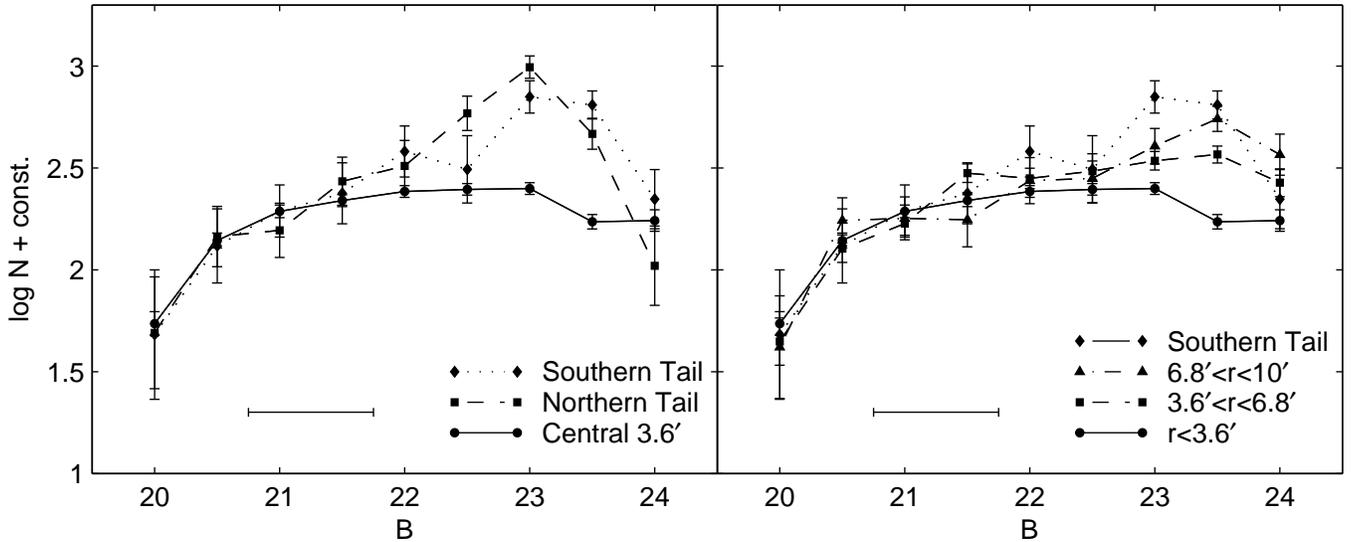}
\caption{Same as Fig. 5, but based on INT data. {\em Left panel}:
Comparison of the cluster center and both tidal tails. {\em Right panel}: LFs in different
annuli around Pal 5's center. For the
ease of comparison, the curves were normalized to fit the LF of the
central region at the bright end.} 
\end{figure*}

\begin{table*}[t]
\begin{center}
\caption{INT Luminosity Functions}
\begin{tabular}{c c r c r c r c r c r} 
\colrule
\colrule
& \multicolumn{2}{c}{Cluster Center} & \multicolumn{2}{c}{Northern Tail} & \multicolumn{2}{c}{Southern Tail} & \multicolumn{2}{c}{$3\farcm 6\le r\le 6\farcm 8$} &
\multicolumn{2}{c}{$6\farcm 8\le r\le 10\arcmin$}\\
B & N & LF & N & LF & N & LF  & N & LF & N & LF\\
\colrule
20.0 & 58   & 54.4  & 10 & 5.6 & 9 & 4.7  &  18  & 16.7  & 6  & 5.2  \\
20.5 & 142  & 139.4 & 25 & 16.6 & 21 & 12.7  &  49  & 47.7 & 23 & 21.6 \\
21.0 & 195  & 193.7 & 24 & 17.8 & 25 & 18.8  & 64  & 63.2  & 23 & 22.1 \\
21.5 & 217  & 218.6 & 50 & 31.1 & 42 & 23.0 & 112 & 111.9 & 24 & 21.8  \\
22.0 & 235  & 242.6 & 70 & 36.9 & 70 & 36.9  &  104 & 105.5 & 37 & 33.9 \\
22.5 & 244  & 248.2 & 113& 67.0 & 77 & 30.1  &  115 & 114.5 & 40 & 34.8 \\
23.0 & 238  & 250.7 & 144& 112.9 & 102 & 68.4  &  124 & 128.7 & 52 & 50.1 \\
23.5 & 149  & 172.2 & 55 & 53.1 & 63 & 62.4 &  120 & 138.3 & 60 & 68.2\\
24.0 & 73   & 174.6 & 5  & 12.0 & 9 & 21.5 &  42  & 100.4 & 19 & 45.4 \\
\colrule
\end{tabular}
\end{center}
\tablecomments{Same as Table 4, but for INT data.}
\end{table*}

For comparison the combined results from both the central
fields in GS01 are also shown in the WFI plot. We chose to normalize each LF
to the curves from the central parts at the bright end using a simple least-squares
algorithm (Piotto, Cool \& King 1997) in the magnitude range from
V=20.5\,mag to V=21.5\,mag (likewise for B), as indicated by a solid bar at the bottom of the
diagrams.  

\subsection{Central region}

Generally, our WFI observations are in good agreement with the V-band LF
from GS01, thus confirming the flattening of the LF towards the faint
end (M$_V\ga4$) in a region located within the core radius. This trend is also 
visible in the INT data (Fig. 6). 
If we confined our data to the same region as covered by the HST in GS01,
we find that our LFs coincide with those from GS01 to within the uncertainties. 
One should
note that the ``faint end'' in our work is far brighter than the
limiting luminosity reached by
GS01, as their LF extends out to M$_V=10.1$.  Yet our
curves drop below the HST-based reference LF in the faintest magnitude
bins of our ground-based observations, leading to an even more
pronounced flattening.  However, it cannot be ruled out that this is
caused by increasing incompleteness or difficulties in accounting for
field contamination in these faint-magnitude bins.  GS01 did not apply
any correction for field star contamination.  Field stars were only
excluded by a hand-drawn main-sequence envelope, similar to our
2$\sigma$ envelope.  Thus there may be remaining contamination of the
main sequence  with non-cluster members in the stellar sample of GS01.  
Another difference is that of the area sampled.  The HST field of view
covered by the GS01 data is $3\farcm 4$ squared 
whereas our LF of the central region
covers $6\farcm 4$ squared, leading to an improved statistics of our
larger sample.

In the brightest bin (V=20\,mag) our scaled number exceeds the
reference curve, a fact that can possibly be attributed to a
preferential saturation of the HST data for brighter stars as well
as to statistical fluctuations because of the small number of stars
at brighter magnitudes.
One distinctive feature of GS01's LF is a dip at M$_V$=5.6\,mag, which
does 
not coincide with our WFI data  
within GS01's statistical
(Poisson) uncertainties.  Considering the larger area covered of our
analysis it appears more likely that the LF flattens constantly
towards the faint end, which is strengthened by the fact that this trend is 
present in data taken with two different instruments. 
This flattening points to a strong deficiency of Pal 5's core in low-mass 
stars.
\subsection{Tidal tails}

After statistical removal of field contamination in the entire fields
B and C, and also in the WFI field F2, there still was a non-zero
number of stars present.  Complementary to the previous selection in
color-magnitude space, these can be statistically ascribed to the
cluster population.  

In contrast to the LF of Pal\,5 itself, the LFs of the tidal
 tails are rather steep compared to the LF
of the cluster center at the faint end, i.e., the tails contain a 
higher fraction of low-mass stars than the central region of the cluster.
The LFs of the tails agree within the
uncertainties indicated by the error bars (left panel of Fig.\ 6). 

The strong
decline in the LFs of both streams at B$\gtrsim$23.5 is due to
systematic effects because of incompleteness effects and field 
contamination and is not believed to be a real levelling off at that
point. 
To test the relevance of outliers and to explore whether the differences
between the northern and southern tail are significant, we applied a 
Kolmogorov-Smirnov (KS) test
the three samples presented in Fig.\ 2.  The test was
truncated at B=23.5\,mag, where incompleteness effects become severe.
We find a probability of 57\% that northern and southern tail are
drawn from the same population.  Thus the results from both northern
and southern stream show consistently a similar, high degree of
enhancement in low-mass stars within the uncertainties.  
On the other hand, the KS
probability is practically zero for the hypothesis that the tails
and the cluster center have the same luminosity distribution.

In order to estimate the degree of depletion or enhancement of 
faint stars in terms of LF flattening, we performed an error-weighted 
least squares fit of a power-law function to our
observed LFs. The LF of the cluster center is the flattest ($x=-0.65\pm0.22$) 
\footnote{The analoguous fit to our WFI and GS01's V-Band data yield
$x=-0.7\pm0.4$ and $x=-1.0\pm0.3$, which underscores the good agreement between those datasets.}
compared 
to those of the northern and southern tail ($x=4.7\pm0.7$ and $x=3.3\pm0.8$) 
at the faint end, i.e., for $21\le$B$\le 23.5$.
These numbers suggest a distinct steepening of the tails'
LF compared to that of the core.
 
This is the first evidence of mass
segregation effects in this cluster: The tails, composed of stars
evaporated from outer regions of the cluster, contain a higher
fraction of faint  
stars. Since the tidal shocks removed at
first stars from these outer regions, there must have been a
pre-existing differentiation in the stellar mass distribution 
between the central part and the
outskirts of the cluster.  This conjecture is 
consistent with the observed 
depletion of faint stars in the core of Pal 5.

\subsection{Mass segregation within Palomar 5} 

In order to look for
further evidence of mass segregation in terms of a radial
variation of the LF, number counts were performed in two additional
regions around the cluster center on our INT data.  These consist of
two annuli each of which has a width of 3$\farcm$2. This width was
chosen in order to ensure that the annuli cover roughly the same
area. The LFs derived for these regions are 
given in Table 5 and shown in the right panel of Fig.\ 6.
There is a gradual rise in slope
with increasing distance from the cluster
center, starting with the strongly flattened LF of the central
cluster population via the annuli's LFs 
toward the low-mass-enhanced function of the tail.
Power-law fits in the 
same magnitude range as used above 
yield indices of $x=1.3\pm0.4$ and $x=2.1\pm0.7$ for $3\farcm 6<r\le 6\farcm 8$, $6\farcm 8<r\le 10\arcmin$, respectively.

A KS-test revealed a probability of $<3\%$ that stars in the
annuli and in the center are drawn from the same 
population.  
Hence we find our conjecture of mass segregation within Pal\,5 
confirmed (which then also permeates to the tails). 
GS01 did not find any evidence of spatially varying 
segregation, but they concentrated on a small region 
(within our central field) in the cluster center itself
and did not have coverage that would have permitted them to
analyze the outer cluster regions.

\section{Mass functions}

In order to be able to unambiguously compare our results, which were obtained 
in different photometric bands, we convert  
luminosities into masses using the M/L relations of Girardi et al.\ (2002, 
2004). These relations offer
the advantage of being available in several photometric systems 
and thus to be well suited for our purposes, and 
reproduce well the observed main sequences in the WFI and INT data.
Isochrones with an age of 11.2\,Gyr and a metallicity of [Fe/H]=$-$1.30 
were chosen, since these values represent Pal 5's actual
characteristics (see Table 1) and yielded satisfactory results.

\begin{figure*}[t]
\includegraphics[angle=0,clip,width=1\hsize]{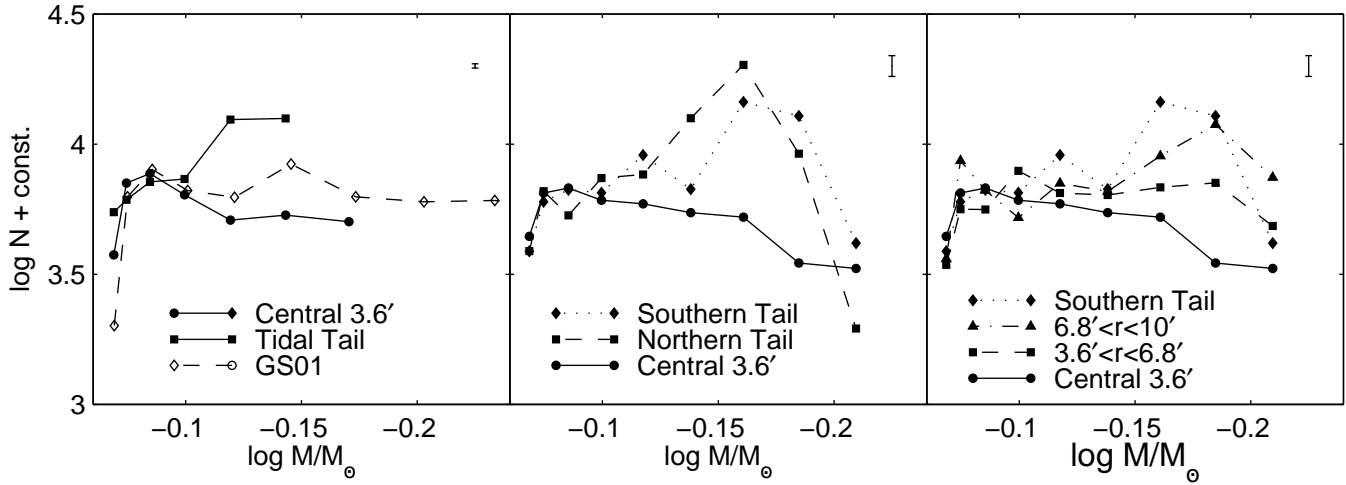}
\caption{Mass Functions for all observed regions in Pal 5, translated from the LFs in Figs. 5, 6 using
M/L relations from Girardi et al. (2002, 2004) for 11.2\,Gyr and [Fe/H]=$-1.30$. The {\em left panel} 
shows the WFI based data, whereas the {\em middle} and {\em right panels} display MFs derived from the INT LFs, each scaled to 
fit the low-mass end of the cluster center. Typical error bars derived from the number counts are
indicated.}
\end{figure*}

In Fig.\ 7 we compare the mass functions (MFs) derived from the V- and 
B-band LFs of all regions as presented in Section 4.  As before,
the curves were normalized to the curve of the cluster center
at the high-mass end.
The magnitude range covered by our observations corresponds to a mass range 
of $M=0.58-0.86\,M_{\sun}$.
Errors in the mass functions were adopted from the LFs and propagated through the
M/L relations. 

Neither one of the curves is particularly well described by a power law over 
the mass range in question. Still, to assess the significance of
the low-mass-star depletion of the cluster core in comparison with the tails,
we determined best-fit power law indices for masses below 
0.82$\,M_{\sun}$. These values are:

\begin{eqnarray*}
x\,&=&\,-0.48 \pm 0.04\mathrm{, \quad GS01}\\
x\,&=&\,-0.63 \pm 0.20\mathrm{, \quad WFI, \,\,r<3\farcm 6} \\
x\,&=&\,-0.63 \pm 0.11\mathrm{, \quad INT,\,\, r<3\farcm 6 } \\
x\,&=&\,\,\,\,\,\,0.77 \pm 0.17\mathrm{, \quad INT,\,\, 3\farcm 6<r\le 6\farcm 8}  \\
x\,&=&\,\,\,\,\,\,0.90 \pm 0.31\mathrm{, \quad INT,\,\, 6\farcm 8<r\le 10\arcmin} \\
x\,&=&\,\,\,\,\,\,0.97 \pm 0.36\mathrm{, \quad INT,\,\, Southern\,\, Tail}\\
x\,&=&\,\,\,\,\,\,1.31 \pm 0.35\mathrm{, \quad INT,\,\, Northern \,\,Tail}\\
x\,&=&\,\,\,\,\,\,1.35 \pm 0.82\mathrm{, \quad WFI,\,\, Tail \,\,(Field\,\, F2)}\\
\end{eqnarray*}
In this notation, $x=1.35$ is the power-law slope of a Salpeter MF.
The MFs from all observations of the central parts of Pal 5 show a high
degree of flattening and good agreement within their uncertainties.
Although our data cover only the higher-mass range of the main sequence, 
the resulting slopes are consistent with 
the estimate of Pal 5 having a MF index no larger than $-0.5$ (GS01).
The slopes of the tidal tails, on the other hand, differ considerably from 
the cluster center, and their formal errors indicate a high degree of 
significance.  It is worth noticing that the MFs of the outermost regions 
in the tidal tails approach the classical Salpeter MF with slopes 
around 1.35.  Similarly, here the values are consistent with Kroupa's
(2001) multiple-part power-law MF of $x = 1.3 \pm 0.3$ in the mass range
of $0.5$ M$_{\odot} < M < 1.0$ M$_{\odot}$. 
The mass segregation between cluster and tails
is well confirmed by the MFs.  

\section{Photometric Binaries} 

Figure 2 (left panels) and Figure 8 show a fairly well-defined sequence
of stars redward of the main sequence of Pal\,5, which we ascribe to
binary stars.  
\begin{figure}[h]
\includegraphics[angle=0,clip,width=8cm]{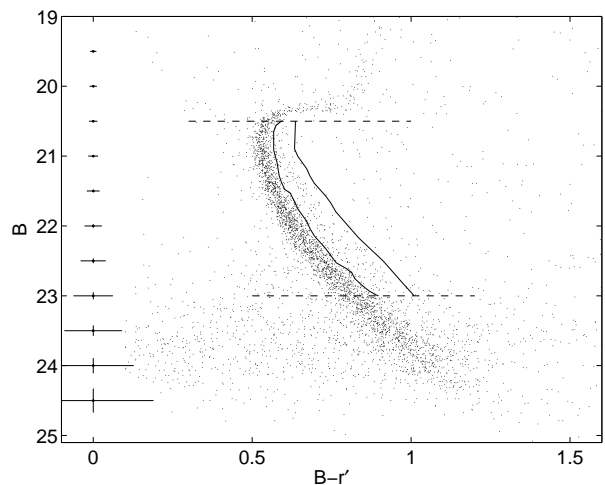}
\caption{CMD of stars within $3\,r_c$. The solid
line is the same 2$\sigma$ envelope as in Fig. 2, shifted by 0.75\,mag
to encompass the binary main sequence. Dashed lines mark the magnitude
limits within which we carried out the analysis of the secondary sequence. 
At the left edge of the diagram typical photometric errors are displayed.}
\end{figure}
In principle, other effects such as rotation (e.g.,
Collins \& Sonneborn 1977; Grebel, Roberts, \& Brandner 1996) can also
account for a color and magnitude shift in the locus of main-sequence 
stars, but these affect primarily high-mass stars.  Rotational
velocities of unevolved stars in globular clusters have been found to
be low (Lucatello \& Gratton 2003).

Only certain types of binaries can be distinguished as ``photometric
binaries'' because of their location in a CMD.  Binaries 
with large mass ratios will be embedded within the primary main 
sequence (Hurley \& Tout 1998; Elson et al.\ 1998).
A visible binary sequence will be composed primarily of
stars with a mass ratio of $q>0.6$ (Pols \& Marinus 1994).  
We will miss wide-separation binaries whose components
are resolved in our photometry, although these should only contribute
sparsely. In the
following, we will call only those stars ``binaries'' whose photometric
properties place them at a location outside of the chosen $2\sigma$-envelope 
around the main sequence. Hence our binaries are at best a subset of the
true number of binaries.

In order to estimate the stellar content of the
binary sequence, we performed number counts similar to those described
in Sect.\ 4. Since the mean locus of the secondary main sequence should
be shifted by no more than 0.75\,mag (unevolved equal-mass binaries)
toward brighter magnitudes, we
chose to shift the 2$\sigma$-envelope by this maximum amount 
and counts of binary candidates were obtained within this shifted envelope.
As the
CMD in Fig. 8 implies, the observed binary sequence can be excellently
accounted for by this
shift, and there is no apparent overdensity near the main sequence
that would correspond to a magnitude shift of more that 0.75\,mag.
We confined our analysis to the
magnitude range $20.5<B<23$, where the secondary main sequence is
defined most clearly. Field star contamination was accounted for in the usual
way via counts in the identical region of the CMD of the comparison field D.

Another source of data points that appear to be photometric binaries is
crowding.  Regular main-sequence stars may be scattered 
away from the primary sequence due to their photometric errors
(see Fig. 8) and pollute the second sequence, hampering the 
determination of the cluster's {\em true} binary content.

Crowding and blends will particularly contribute to these spurious
binaries (Walker 1999).  On the other hand, Pal\,5 is a very loose,
sparse cluster with little crowding, so one would expect less of an
effect than in more typical, rich, dense globular clusters. 

In order to test the possibility of photometric scatter as
the major source of the redward spread in the CMD, we divided our data
into two samples following the procedure employed by Walker (1999):
``Sample 1'' was purged of stars with standard errors (as determined by
DAOPHOT) larger than two median standard errors of all stars within a
given magnitude bin of 1\,mag width. Yet this procedure only removed a few
outliers and is thus basically identical to the inclusion of all stars.
Our ``sample 2'' contains only those stars within two standard deviations
of the mean of the photometric errors.
Additionally, the data were cut off
at standard errors greater than 0.02\,mag, thus forming a low-error
``sample 3''.
This cut basically removes stars with B\,$\ga 22.5$\,mag.
For all samples, we determined the distribution of colors around the
main sequence fiducial.  As the left panel of Fig.\ 9 shows, {\em neither}\
of the distributions is symmetric across the peak.  Each distribution shows
a clear excess of stars with redder colors.  The difference
between the three samples is small, and the samples 1 and 2 are almost
identical.  Sample 3 exhibits a narrower color
distribution both blueward and redward of the fiducial.
This behavior is different from what
was found by Walker (1999) for the rich globular cluster NGC 2808.

The same procedure was carried out on our artificial star sample, which is
shown in
the right panel of Fig.\ 9. Also these distributions are asymmetric and
all show an excess in numbers toward redder colors.  Again, the samples
1 and 2 are essentially indistinguishable.  Only sample 3 shows a slightly
reduced red excess.

\begin{figure}[h]
\includegraphics[angle=0,clip,width=8.1cm]{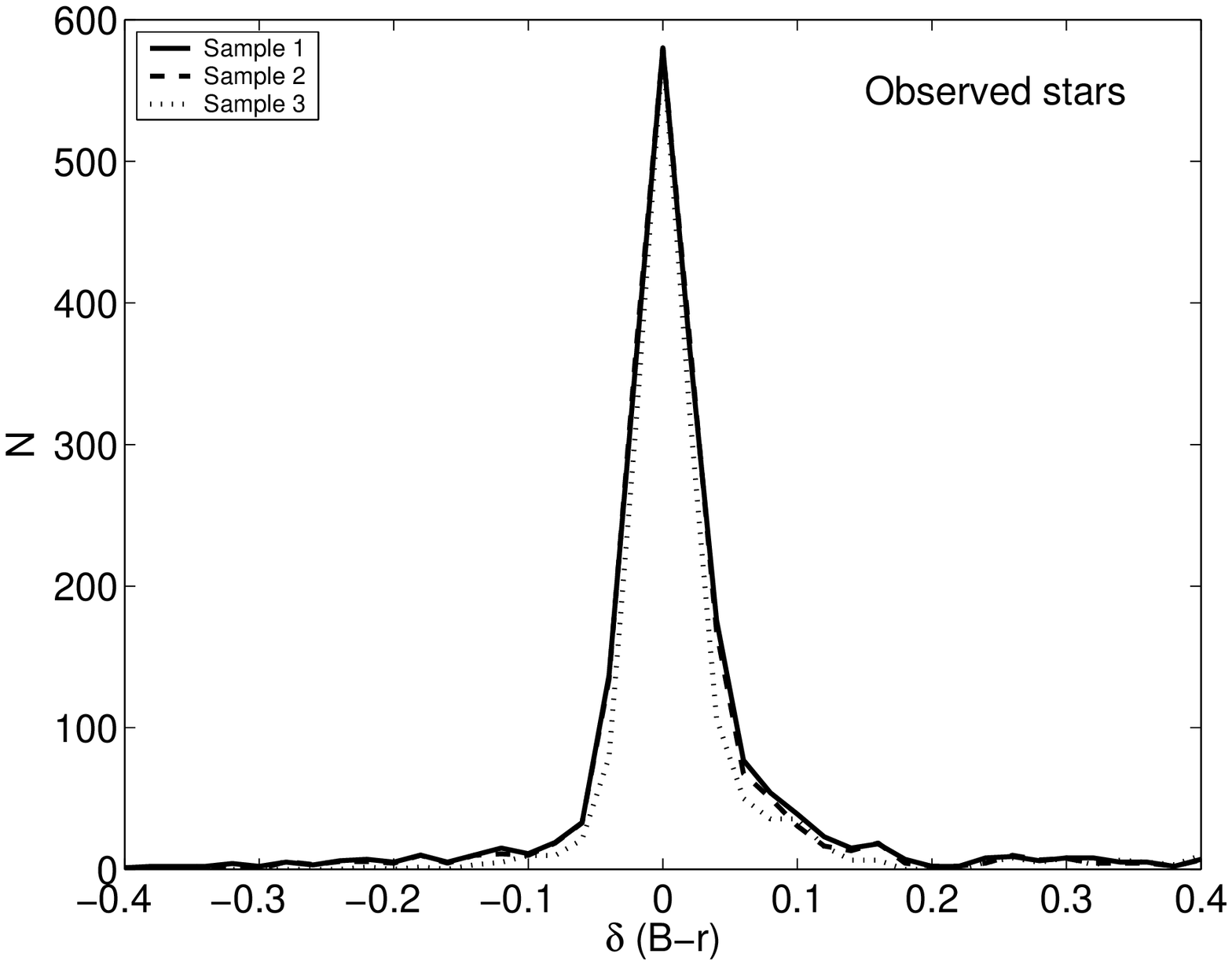}
\includegraphics[angle=0,clip,width=8.1cm]{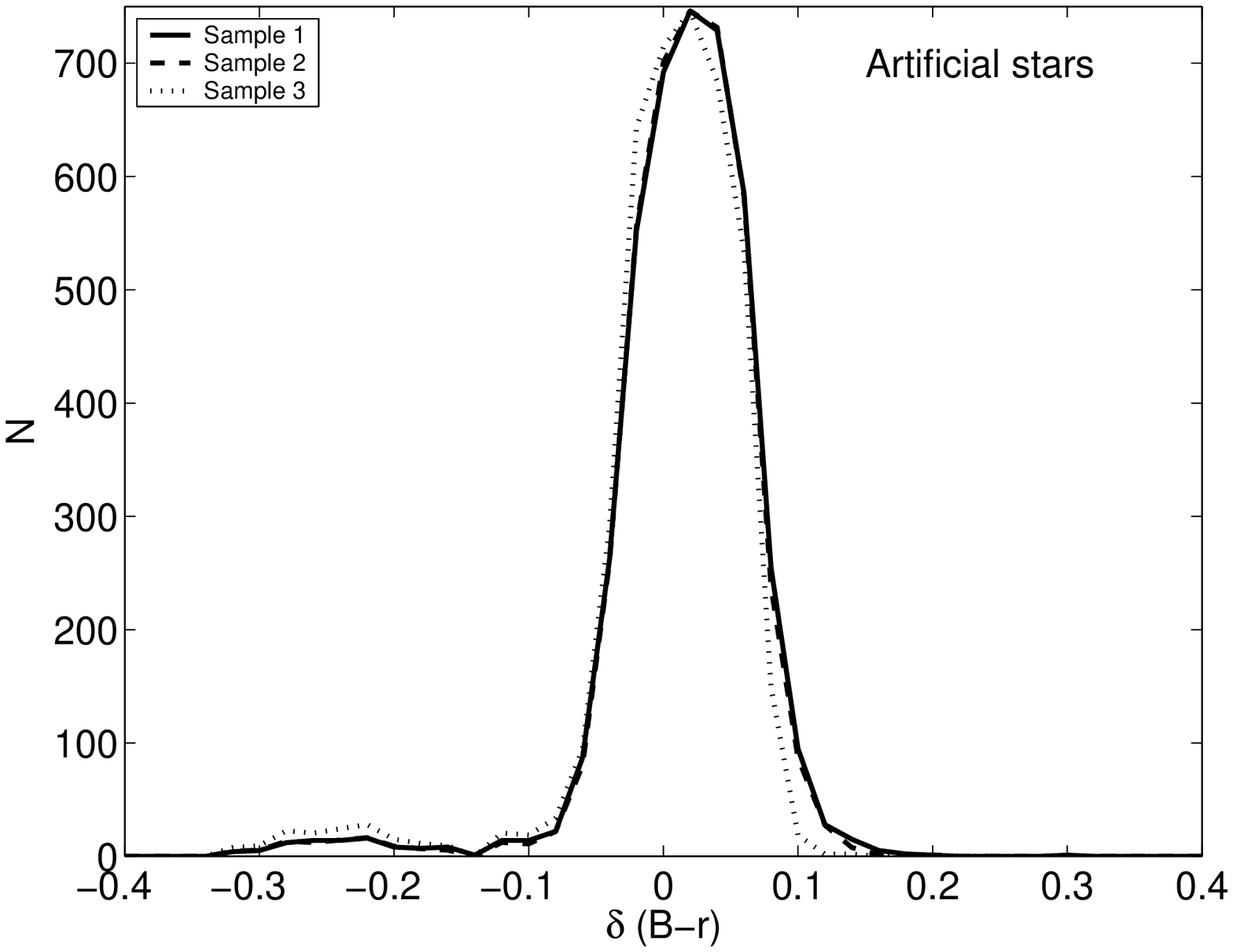}
\caption{Histograms of the colour distribution of stars 
around the MS fiducial (normalized to the peak values). On the {\em left} side, observed stars from our
INT data are shown. The three samples were drawn from stars within 3 core radii
according to cut offs in their photometric errors. The {\em right} panel displays the respective 
results from our artificial star experiments.}
\end{figure}

We conclude that photometric scatter alone cannot
explain the secondary sequence seen in the CMD,
since also the observed low-error samples unequivocally exhibit such a
feature.  Furthermore, the differences between sample 2 and 3 indicate
that photometric scatter becomes mainly significant at
B\,$\ga 22.5$\,mag.
 
It is worth noticing, however, that the color distribution of the artificial
stars is generally more dispersed towards the red: Its average of 0.015\,mag (with a 1\,$\sigma$-width 
of 0.04\,mag) compares to an average of practically zero (and a width of 0.02\,mag) of
the distribution of INT-observed stars. One reason for this shift is the high occurrence of red field stars 
(see our CMDs, Fig. 3), resulting in a larger probability that an artificial star will blend with
such a red star (Aparicio \& Gallart 1995). This will yield a redder color of the recovered star.

\subsection{Radial variations in binary fractions versus single stars?}

We define the binary fraction $f_b$ as the
field- and completeness-corrected number
ratio of binary candidates 
within the shifted $2\sigma-$envelope
to the combined number of binaries and {\it prima facie}\ 
single stars (primary sequence within the original 2$\sigma$-region).  
In the cluster center, i.e., for $r<r_c,\, f_b$ is found to be
(9 $\pm$ 1)\%, with the uncertainty being purely based on 
(Poisson-) uncertainties in
number counts.  This is a value in good agreement with typical globular cluster
binary frequencies of 3--30\% (see reviews of Hut et
al.\ 1992; Meylan \& Heggie 1997, and references therein). There is now 
strong evidence that most globular 
clusters began their lives with a significant primordial binary content 
of at least 10\% (McMillan, Pryor \& Phinney 1998).
Moreover, low density globular clusters 
suffering from extensive tidal mass loss have been
shown to maintain a fraction of soft binaries at the high end of this observed
range (Yan \& Cohen 1996; McMillan et al. 1998).  Under
elaborate assumptions about the binary mass spectrum and orbits, Odenkirchen
et al.\ (2002) simulated the behavior of binaries in Pal\,5 and
estimated $f_b$ to be roughly (40 $\pm$ 20)\%, which is significantly
higher than our observations imply.  This may be due to our insensitivity to
binaries that are not near-equal-mass pairs. Hence, a concise
determination of binary fractions necessitates allowance for their
mass ratios or the use of statistical methods (Bellazzini et al.\
2002a), but this is beyond the scope of our rough photometric 
estimate. 

To determine radial variations in the occurrence of binaries, the
secondary sequence was analyzed out to 3 core radii. The resulting
values for $f_b$ are illustrated in Fig. 10. 

\begin{figure}[h]
\includegraphics[angle=0,clip,width=8cm]{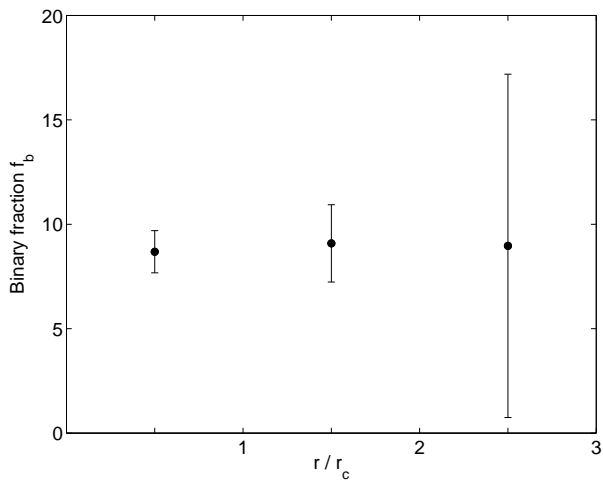}
\caption{Estimated binary fraction as measured
in concentric annuli around the center of Pal\,5: $r\le r_c$, $r_c<r\le 2\,r_c$ and $2\,r_c<r\le 3\,r_c$.  
Also indicated are Poisson error bars.}

\end{figure}

In the outer parts of
the cluster, $f_b$ is (9 $\pm$ 8)\% for $2r_c<r<3r_c$, hence the 
uncertainties here are too large to permit a meaningful measurement.
Within two core radii, the binary fraction remains roughly constant as a 
function of radius within the uncertainties.
Based on these data we do not find evidence of mass segregation in
the binary component as a function of radius in the sense that more
binaries are residing in the central region.  

\subsection{Radial variations in binary masses?}

Finally, we consider the question whether more massive binary systems
differ in their radial distribution from that of less massive ones. 
To distinguish between the two, we adopt a simple luminosity cut and
then measure the ratio
N$_b(20.5 < B < 22)$/N$_b( 22 < B < 23)$ within different radial bins.  
These cuts correspond to approximate mass ranges of 
$0.76-0.84\,M_{\odot}$, $0.69-0.76\,M_{\odot}$, respectively.
Between two and three core radii, number statistics are too poor.
In the annulus within one core radius, we find a ratio of $1.00\pm 0.25$ 
and in the annulus ranging from
two to three core radii, this ratio is $0.55\pm 0.25$.
This compares to 1.12$\pm$ 0.07 and 1.01$\pm$ 0.11 in the same two annuli
for stars on the primary main sequence.
While the ratios do not vary much 
along the ``single''-star main sequence, the binary
ratios may indicate a tendency for more massive binaries to be more
centrally concentrated (although the uncertainties are large).

\subsection{Binaries: primordiality versus dynamical evolution}

The origin of the binaries can be explored by comparing rates of the
processes that produce binaries (Bellazzini et al.\ 2002a).  
Tidal captures should only have contributed marginally during Pal 5's
evolution, regardless of the environment in which the cluster resided
(i.e., at apocenter in the halo or during disk shocks). The latter
becomes obvious when inserting Pal~5's low velocity dispersion $\sigma$ 
and respective estimates for the stellar
density $\nu$ into the formula by Lee \& Ostriker (1986) 
\begin{displaymath}
t_{tid.capt.}\,\approx 10^{12}\,yr\,\times\,\left(10^5\,pc^{-3}/\nu\right)\,\left(\sigma/100\,km\,s^{-1}\right)^{1.2}
\end{displaymath}
(after  Binney \& Tremaine 1994)
-- both assuming the current value (Odenkirchen et al.\ 2002, 2003)
and allowing for a variation of an order of magnitude. The resulting capture
rates are less than 1 over Pal~5's present age.
The rate of binary systems forming by three-body interactions is
estimated to be $\sim 0.1/(N\ln N)$ per relaxation time (Binney
\& Tremaine 1994).  Using Pal~5's total present-day 
mass of $\sim 5000$\,$M_{\sun}$
(Odenkirchen et al.\ 2002) and an average stellar mass of
1\,$M_{\sun}$, this corresponds to $\sim 10^{-7}$ during the cluster's
lifetime.  Hence, the most likely origin of Palomar 5's binary
population is primordiality {\it if the present properties of Pal\,5
are representative of its past mass and density, and if the cluster
is in dynamical equilibrium}.

Dehnen et al.\ (2004) argue that the large size and
low concentration of Pal\,5's main body are due to expansion after
the last disk shock.  They note that Pal\,5 is at least two times larger
than its theoretical tidal radius.  The internal dynamical time scales
of Pal\,5 exceed the time between disk passages, so that the external 
tidal shocks dominate the dynamical evolution.  Pal\,5 {\em may} have been
already a low-concentration, low-mass cluster to begin with, which
together with its highly eccentric orbit would have facilitated its 
dissolution.
Extrapolating back, Dehnen et al.\ (2004) estimate Pal\,5's original mass
to have been $\le 70,000$~M$_{\odot}$.  The estimates 
provided by these authors again suggest long dynamical time scales
with a two-body relaxation time of $\sim 20$ Gyr.  This, too, then supports
the assumption that the binaries in Pal\,5 and its mass
segregation are largely primordial.   On the other hand, Dehnen et al.'s\ (2004)
N-body simulations only cover a few Gyr, and they caution that 
extrapolating back further may be dangerous because of the unknown
changes in the Galactic disk potential.  Hence it cannot be excluded
that Pal\,5 was initially significantly denser and did experience
dynamical mass segregation.  

\section{Comparison with other clusters}

We have shown that Pal 5's main sequence LF is significantly deficient in faint
stars compared to the LF of its tidal tails and that its mass
function thus is strongly flattened. Now the question arises how this
behavior compares to other globular clusters.  Or in other words: how 
strong is the
depletion of the cluster itself compared to the LF or MF of other clusters?
Table 6 lists four different clusters that will be used for an
illustrative comparison.

\begin{table*}[t]
\begin{center}
\caption{Comparison Clusters}
\begin{tabular}{c c c c c }
\colrule
\colrule
Object & [Fe/H] & concentration & $\log\,t_{rh}$ & Reference \\
\colrule
Pal\,5   & $-$1.43 & 0.66 & 9.89 & cf. Table 1 \\
M\,10    & $-$1.52 & 1.40 & 8.86 & Piotto \& Zoccali (1999) \\
M\,22    & $-$1.64 & 1.31 & 9.22 & Albrow, DeMarchi \& Sahu (2002)\\
NGC\,288 & $-$1.39 & 0.96 & 9.26 & Bellazzini et al. (2002b)\\
NGC\,6397& $-$1.82 & 2.50 & 8.46 & Andreuzzi et al. (2004)\\
\colrule
\end{tabular}
\end{center}
\tablecomments{Parameters of Pal 5 and globular clusters
with different characteristics. Values that are not stated in the 
respective references were taken from the Harris (2003) database.}  
\end{table*}  

All of these globular clusters were found to
exhibit hints of flattening at the faint end of their overall LF, which
becomes manifest in the comparison to a whole sample of globular
clusters (e.g.,  Piotto
\& Zoccali 1999). Likewise, power-law indices of many cluster MFs
tend to be fairly shallow as the comparison of the LFs of a few GCs
 implies (see Fig. 11).

\begin{figure}[t]
\includegraphics[angle=0,clip,width=8.6cm]{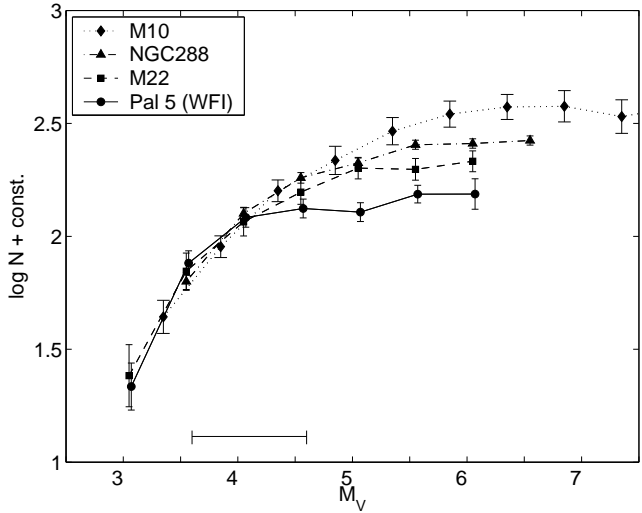}
\caption{Luminosity functions of Pal 5's
center (WFI data), M\,10 (Piotto \& Zoccali 1999), M\,22 (Albrow,
DeMarchi \& Sahu 2002) and NGC\,288 (Bellazzini et al.\ 2002b).
M\,10's LF can be described best by a mass function with a power-law
exponent $x=0.5$, whereas Pal 5's LF is significantly flatter ($x\approx $--0.6).}
\end{figure}

Among these, M\,10 is the cluster with the steepest LF (an index of $x\approx +0.5$ 
fits this cluster's MF). Leon, Meylan \&
Combes (2000) found weak evidence for a tidal extension resulting from
a recent disk crossing. They predict mass segregation, although there
is no observational evidence for this effect so far (Hurley, Richer \& Fahlman 1989).  
The differences
to other clusters' LFs are attributed to primordial differences,
i.e., different IMFs, 
or to a moderate internal dynamical evolution, as 
reflected in its short relaxation time (Piotto \& Zoccali 1999).
The LFs of M\,22 and NGC\,288 are, on the other hand, distinctly
flattened compared to M\,10. In both these clusters, mass segregation
has been established and it is known for NGC\,288 that it exhibits a
highly inclined and eccentric orbit making it sensitive to tidal
shocks during passages close to the Galactic center (see reference in
Table 6). It then appears even more remarkable that Pal 5's LF falls
below the already significantly depleted LFs of the other potentially
tidally pruned clusters (although similar tidal features as around
Pal\,5 have yet to be detected in these clusters). 
This may support the suggestion by Dehnen et al.\
(2004) that in order to be so significantly disrupted, a cluster must
have been of comparatively low mass and low concentration {\it ab 
initio}, whereas M22 and NGC\,288 are both more compact and more massive.

The presence or absence of mass segregation is also demonstrated
in Fig. 12, where we compare LFs of the globular clusters
from Table 6 that were obtained at different
distances from each cluster's center. 
\begin{figure*}[t]
\includegraphics[angle=0,clip,width=1\hsize]{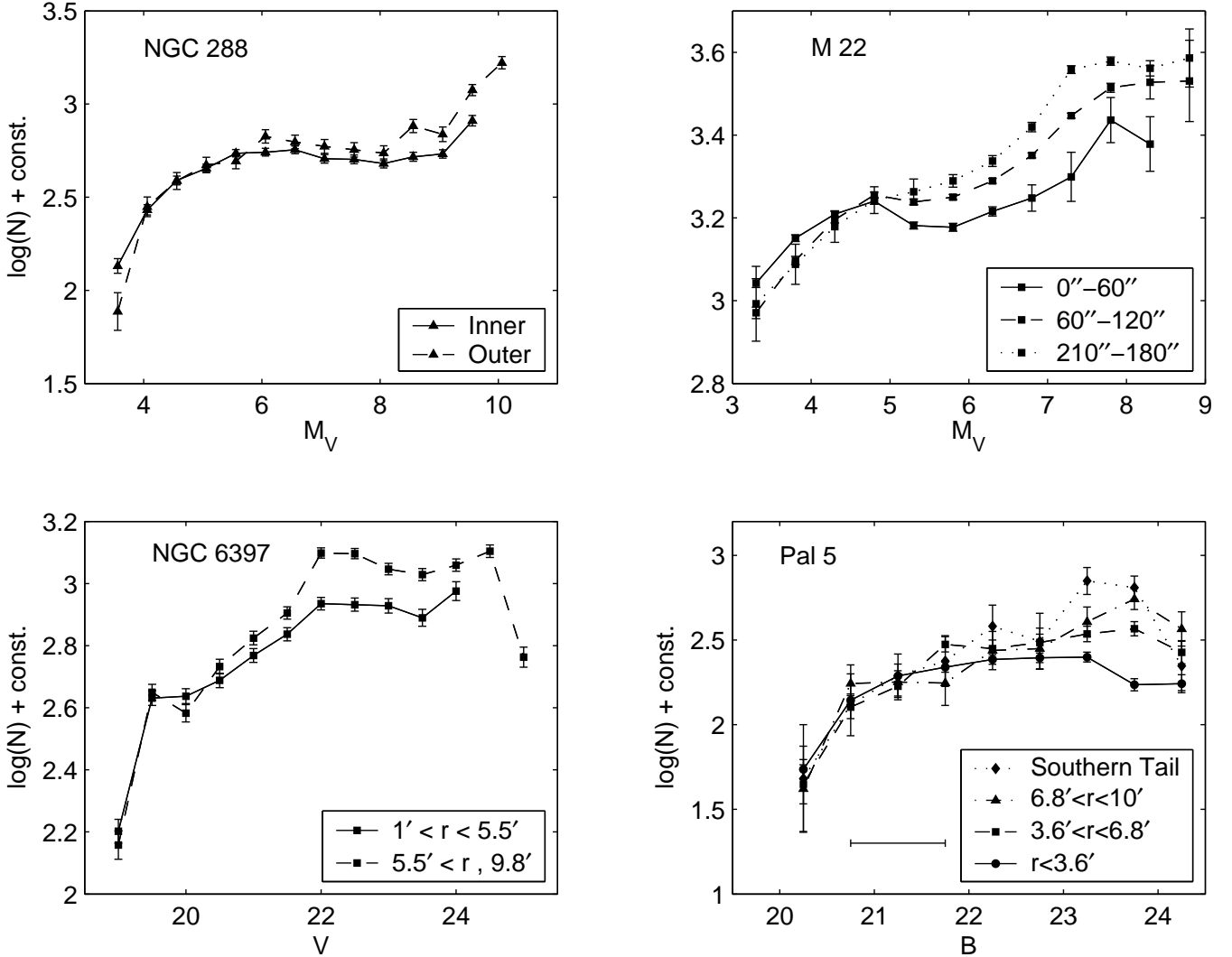}
\caption{Spatially resolved LFs of the clusters
in Table 6, each scaled such that their number counts match one another
at their bright end.  The {\em
inner} and {\em outer} curves of NGC\,288 refer to a field
near its half-mass radius (``inner'') 
and a region located at a larger distance from the cluster center. 
M\,22's LF was analyzed  in annuli around the
center ($r_c\,=\,1\farcm 4$). 
The data are from Bellazzini et al.\ (2002b) for
NGC\,288, Albrow, DeMarchi \& Sahu (2002) for M\,22 and Andreuzzi et al. (2004) for NGC\,6397. Pal 5's LF is
the INT-based LF derived in the previous chapters.}
\end{figure*}
A comparison of two LFs from HST observations of fields
near two and three core radii in NGC\,288 (top left) reveals a
significant deficiency of faint stars in the inner field compared to
the outer one, thus confirming the occurrence of mass segregation,
which is in good agreement with the theoretical predictions
(Bellazzini et al. 2002b).  
M\,22's LF (top right panel)
has been analyzed in annuli reaching out to five core radii (Albrow,
DeMarchi \& Sahu 2002). There is a significant enhancement of massive
stars in the core compared to the regions outside the core. As the
authors show by numerical modelling, the high degree of observed mass
segregation can be entirely explained by standard relaxation processes
within the cluster. 
Various observations (Andreuzzi et al. 2004; De Marchi, Paresce \& Pulone 2000; 
Piotto \& Zoccali 1999) have shown that the core collapsed 
metal poor cluster NGC\,6397, which is also the Galactic GC closest to us, 
has a remarkably flattened MF.
Estimates of a power-law yielded $x=-0.5$ for the present-day MF, placing it at the lower end of the
MF scale. 
The deficit in low-mass stars in this cluster is attributed to 
evaporation and tidal shocking, both effects operating simultaneously.
Its vulnerability to tidal shocks is underscored by the cluster's highly oscillating orbit (Dauphole et al. 1996), 
bringing it close to the dense regions of the Galacit plane near the center.

In addition to this high degree of depletion, analyses of the MFs of NGC\,6397
at different radii have revealed a significant difference in the MF slope,
where this mass segregation indicates that NGC\,6397 is a dynamically 
relaxed system (see its short relaxation time, Table 6).

Hence mass segregation is a common phenomenon in globular clusters.
Depending on the intrinsic cluster properties and on their orbits, 
it may be primordial or evolutionary; most likely a mixture of both.
  
\section{Summary}

The halo globular cluster Pal\,5 is the first globular cluster around
which well-defined, extended tidal tails were detected.  The cluster
is believed to be in the final stages of total dissolution and stands
out due to its low mass, low density, and comparatively large angular
extent.  If Pal\,5 was a low-concentration cluster to begin with, one
would expect that internal dynamical effects should have had little
impact on its evolution owing to the long stellar encounter time scales.
Under these assumptions, one may also expect that mass segregation within
the cluster would be small unless primordial mass segregation took place.
External tidal effects due to disk shocks should dominate.  Consequently
one might expect that the mass spectrum in Pal\,5's tails would closely 
resemble the mass spectrum in the cluster itself.

In the present paper LFs for different regions in the
globular cluster Pal 5 were measured.  The LF of the cluster's
central region was found to be in good agreement with that previously
published by Grillmair \& Smith (2001) based on HST data.  
Comparison with other globular clusters shows that Pal
5's central LF is significantly depleted in low mass stars. 

We measured LFs also in various annuli around the cluster center
as well as in the tidal tails of Pal\,5. 
While the cluster center is depleted in low-mass stars, 
the LFs become increasingly steep with increasing distance
from the cluster center, and the tidal tails are enhanced in
faint, low-mass stars.  

This trend is also to be seen in the mass functions that were derived 
from the observed LFs. With a power-law exponent of $x\approx -0.6$ Pal\,5
itself is at the lower limit of globular cluster MFs.
This high degree of flattening is unusual for globular clusters.  
Likewise, the tidal tails' MFs differ significantly from that of the core and 
are similar to Salpeter's (1955) or Kroupa's (2001) power-law MFs.

Hence there is clear evidence of 
a radially varying mass spectrum and thus segregation. 

Mass segregation is not obvious when investigating the distribution
of photometric 
binary stars identified via a binary main sequence in color-magnitude
space in comparison to the primary, single-star main sequence.  However, 
more massive binary system candidates are more strongly concentrated 
toward the cluster center than less massive ones.  

The discovery of mass segregation does not appear to be 
self-evident, since the time scales for {\em dynamical} mass segregation
for low-concentration clusters are of the order of more than a Hubble time.  
On the other hand,
mass segregation has been shown to occur in open clusters (e.g., Bonatto \&
Bica 2003 and references therein).  In the case of Pal\,5, it is difficult
to assess how much of the observed mass segregation might be primordial
and how much might be due to evolutionary effects.  While we can fairly 
accurately determine Pal\,5's present-day mass, structural parameters, 
and kinematics, it remains unclear what its initial conditions were.  
N-body simulations by Dehnen et al.\ (2004) together with the mass loss
rates derived by Odenkirchen et al.\ (2003) 
suggest that Pal\,5 once was more than 10 times as massive than
observed today (present-day mass: $\sim 5000$ M$_{\odot}$).
It may have been a low-density cluster to begin with, possibly with a
certain amount of primordial mass segregation.  As lower mass stars would
then have had a more extended spatial 
distribution, they would also have been more easily stripped from the
cluster during disk passages.  The effect of external tidal forces 
on cluster stars increases with their distance from the cluster center
proportionally to $r^3$.  The internal dynamics will certainly
have been affected by external tides as those led to structural changes
in the cluster as a whole (in particular, expansion and oscillations).
Possibly external effects may even have sped up mass segregation within
the cluster.  
Detailed N-body simulations with star particles covering a range of 
masses are required to explore the interplay of these effects in
more detail.

\acknowledgments

We thank Ortwin Gerhard for valuable discussions.
AK and EKG gratefully acknowledge support by the Swiss National Science 
Foundation through grant 200021-101924/1.
This research has made extensive use of NASA's Astrophysics Data System.


\begin{thebibliography}{999}
%
\bibitem[Aguilar, Hut, \& Ostriker(1988)]{aguilar1988} Aguilar, L., Hut, P., \& Ostriker, J.~P.\ 1988, \apj, 335, 720 
%
\bibitem{albrow02} Albrow, M.D., DeMarchi, G., \& Sahu, K.C 2002, ApJ, 579, 660
%
\bibitem{Aparicio95}  Aparicio, A., \& Gallart, C. 1995, AJ, 110, 2105
%
\bibitem{andreuzzi01} Andreuzzi, G., De Marchi, G., Ferraro, F.R., Paresce, F., Pulone, L., \& Buonanno,R.  2001, A\&A, 372, 851
%
\bibitem{andreuzzi04} Andreuzzi, G., Testa, V., Marconi, G., Alcaino, G., Alvarado, F., \& Buonanno, R. 2004, A\&A subm. (astro-ph/0406309) 
%
\bibitem{bellazzini02a} Bellazzini M., Fusi Pecci, F., Messineo, M., Monaco, L., \& Rood, R.T. 2002a, AJ, 123, 1509
%
\bibitem{bellazzini02b} Bellazzini M., Fusi Pecci, F., Montegriffo, P., Messineo, M., Monaco, L., \& Rood, R.T. 2002b, AJ, 123, 2541
%
\bibitem{Binney94} Binney, J., \& Tremaine, S. 1994 Galactic Dynamics, Princeton University Press
%
\bibitem{Bolte04} Bolte, M. 1989, ApJ, 341,168
%
\bibitem[Bonatto \& Bica(2003)]{Bona2003} Bonatto, C.~\& Bica, E.\ 2003, \aap, 405, 525 
%
\bibitem[Chabrier \& M\'era(1997)]{chamer1997} Chabrier, G.~\& M\'era, D.\ 1997, \aap, 328, 83 
%
\bibitem[Collins \& Sonneborn(1977)]{Coll1977} Collins, G.~W.~\& Sonneborn, G.~H.\ 1977, \apjs, 34, 41
%
\bibitem[1978]{cousins} Cousins, A.W.J. 1978, MNASSA, 37, 8
%
\bibitem{Dauphole96} Dauphole, B., Geffert, M., Colin, J., Ducourant, C., Odenkirchen, M., \& Tucholke, H.-J. 1996 A\&A, 313, 119
%
\bibitem[Dehnen et al.(2004)]{Dehnen2004} Dehnen, W., Odenkirchen, M., Grebel, E.K., \& Rix, H.-W. 2004, AJ, 127, 2753
%
\bibitem{DeMarchi00} De Marchi, G., Paresce, F., \& Pulone L. 2000, ApJ, 530, 342  
%
\bibitem{Elson} Elson, R.A.W., Sigurdsson, S., Davies, M., Hurley, J., \& Gilmore, G. 1998, MNRAS, 300, 857  
%
\bibitem[Girardi et al.(2002)]{Gira2002} Girardi, L., Bertelli, G., Bressan, A., Chiosi, C., Groenewegen, M.~A.~T., Marigo, P., Salasnich, B., \& Weiss, A.\ 2002, \aap, 391, 195 
%
\bibitem[Girardi et al.(2004)]{Gira2004} Girardi, L., Grebel, E.K., Odenkirchen, M., \& Chiosi, C. 2004, A\&A, 422, 205
%
\bibitem[Grebel(2001)]{Greb2001RvMA...14..223G} Grebel, E.~K.\ 2001, Reviews of Modern Astronomy, 14, 223
%
\bibitem[Grebel, Roberts, \& Brandner(1996)]{Gre1996} Grebel, E.~K., 
   Roberts, W.~J., \& Brandner, W.\ 1996, \aap, 311, 470 
%
\bibitem{grillm01} Grillmair, C.J., \& Smith, G.H. 2001, AJ, 122, 3231
%
\bibitem[Gunn et al.(1998)]{Gunn1998} Gunn, J.~E., et al.\ 1998, 
   \aj, 116, 3040
%
\bibitem{harr96} Harris, W.E. 1996, AJ, 112, 1487 
%
\bibitem{harr99} Harris, W.E. 2003, {\tt http://www.physics.mcmaster.ca/resources/\\  globular.html}
%
\bibitem{Hurley98} Hurley, J., \& Tout, C.A. 1998, MNRAS, 300, 977
%
\bibitem[Hurley, Richer, \& Fahlman(1989)]{Hurl1989} Hurley, 
    D.~J.~C., Richer, H.~B., \& Fahlman, G.~G.\ 1989, \aj, 98, 2124 
%
\bibitem{hut92} Hut, P. et al. 1992, PASP, 104, 981
%
\bibitem{johnson53} Johnson, H.L., \& Morgan, W.W. 1953, ApJ, 117, 313
%
\bibitem[King, Sosin, \& Cool(1995)]{King1995} King, I.~R., Sosin, C., \& Cool, A.~M.\ 1995, \apjl, 452, L33 
%
\bibitem{Koch2004} Koch, A., Odenkirchen, M., Grebel, E.K., \& Caldwell, J.A.R. 
	2004, AN, 325, 299
%
\bibitem[Kroupa(2001)]{kroupa2001} Kroupa, P.\ 2001, \mnras, 322, 231 
%
\bibitem{Lee86} Lee, H.M., \& Ostriker, J.P. 1986, ApJ, 310, 176
%
\bibitem[Lee, Fahlman, \& Richer(1991)]{lee1991} Lee, H.~M., Fahlman, G.~G., \& Richer, H.~B.\ 1991, \apj, 366, 455
%
\bibitem{leon00} Leon, S., Meylan, G., \& Combes, F. 2000, A\&A, 359, 907
%
\bibitem[Lucatello \& Gratton(2003)]{Luca2003} Lucatello, S.~\& Gratton, 
   R.~G.\ 2003, \aap, 406, 691 
%
\bibitem{manfroid01} Manfroid, J., Selman, F., \& Jones 2001, ESO Messenger 104, 16
 %
\bibitem{martell02} Martell, S.L., Smith, G.H., \& Grillmair, C.J. 2002,  BAAS, 201, 711
%
\bibitem{McMillan98} McMillan, S.L.W., Pryor, C., \& Phinney, E.S. 1998, Highlights in Astronomy, 11, 616 
%
\bibitem{meylan97} Meylan, G., \& Heggie, D.C. 1997, A\&ARv, 8, 1 
%
\bibitem[Murray \& Lin(1996)]{murlin1996} Murray, S.~D.~\& Lin, D.~N.~C.\ 1996, \apj, 467, 728 
%
\bibitem[Odenkirchen et al.(2001)]{odenk2001} Odenkirchen, M.~et al.\ 2001, \apjl, 548, L165 
%
\bibitem[Odenkirchen et al.(2002)]{odenk2002} Odenkirchen, M., Grebel, E.~K., Dehnen, W., Rix, H., \& 
	Cudworth, K.~M.\ 2002, \aj, 124, 1497 
%
\bibitem[Odenkirchen et al.(2003)]{odenk2003} Odenkirchen, M.~et al.\ 2003, \aj, 126, 2385 
%
\bibitem[Oh \& Lin(1992)]{oh1992} Oh, K.~S.~\& Lin, D.~N.~C.\ 1992, \apj, 386, 519 
%
\bibitem{paresce00} Paresce, F., \& De Marchi, G. 2000 , ApJ, 534, 870 
%
\bibitem{piotto97} Piotto, G., Cool, A.M., \& King, I.R. 1997, AJ, 113, 1345
%
\bibitem[Piotto \& Zoccali(1999)]{piozoc1999} Piotto, G.~\& Zoccali, M.\ 1999, \aap, 345, 485 
%
\bibitem[Pols \& Marinus(1994)]{pols1994} Pols, O.R., \& Marinus, M. 1994, A\&A, 288, 475
%
\bibitem[Salpeter(1955)]{salpeter1955} Salpeter, E.~E.\ 1955, \apj, 121, 161 
%
\bibitem{schech93} Schechter, P.L., Mateo, M., \& Saha, A. 1993, PASP, 105, 1342
%
\bibitem{smith86} Smith, G.H., McClure, R.D., Stetson, P.D., Hesser, J.E., \& Bell, R.A. 1986, AJ, 91, 842
%
\bibitem{smith02} Smith, J.A. et al. 2002, AJ, 123, 2121
%
\bibitem[Stetson(1987)]{Stet1987} Stetson, P.~B.\ 1987, \pasp, 99, 191 
%
\bibitem{Stetson1994} Stetson, P.B. 1994, PASP, 106, 250
%
\bibitem{SDSS} Stoughton, C. et al. 2002, AJ, 123, 485
%
\bibitem{Walker99} Walker, A. 1999, AJ, 118, 432 
%
\bibitem{Yan96} Yan, L., \& Cohen, J.G. 1996, AJ, 112, 1489
%
\end{thebibliography}
\end{document}